\documentclass[aps,prd,superscriptaddress,showpacs]{revtex4}
\usepackage{graphicx}
\usepackage{dcolumn}
\usepackage{bm}
\usepackage{natbib}
\usepackage{multirow}

\newcommand{\beq}{\begin{equation}}
\newcommand{\eeq}{\end{equation}}
\newcommand{\beqa}{\begin{eqnarray}}
\newcommand{\eeqa}{\end{eqnarray}}

\def\affilmrk#1{$^{#1}$}
\def\affilmk#1#2{$^{#1}$#2;}

\def\bkl{2}

\begin{document}

\title{Extracting primordial non-gaussianity without cosmic variance}  
\author{
Uro\v s Seljak \affilmrk{\bkl,}
}
\address{
%{\it
\parshape 1 -1cm 20cm
%\parbox[l]{18cm}{
\affilmk{}{Institute for Theoretical Physics, University of Zurich, Zurich, Switzerland} \\
\affilmk{\bkl}{Physics Department and Lawrence Berkeley National Laboratory, University of California, Berkeley, California 94720, USA}
}
\date{\today}

\begin{abstract}

Non-gaussianity in the initial conditions of the universe is one of the 
most powerful mechanisms to discriminate among the competing theories of the early universe. 
Measurements using bispectrum of cosmic microwave background anisotropies are limited by the 
cosmic variance, i.e. available number of modes. 
Recent work has emphasized the possibility to probe non-gaussianity of local type using the scale dependence 
of large scale bias from highly biased tracers of large scale structure. However, this power spectrum method is also limited 
by cosmic variance, finite number of structures on the largest scales,
and by the partial degeneracy with other cosmological parameters that can mimic the same effect. 
Here we propose an alternative method that solves both of these problems. It is based on the idea that on large scales halos are biased, but not stochastic, 
tracers of dark matter: by correlating a highly biased tracer 
of large scale structure against an unbiased tracer one eliminates the cosmic variance error, which can lead to a high 
signal to noise even from the structures comparable to the size of the survey.  
The error improvement on non-gaussianity parameter $f_{nl}$ relative to the power spectrum method 
scales as $(P\bar{n}/2)^{1/2}$, where $P$ and $\bar{n}$ is the 
power spectrum and the number density of the biased tracer, respectively. For 
luminous red galaxies at $z\sim 0.5$ we get an error improvement 
of about a factor of three over the current constraints, assuming unbiased tracers are 
also available with a high number density. For an ideal survey out to $z \sim 2$ the corresponding 
error reduction can be as large as a factor of seven, 
which should guarantee a detection of non-gaussianity from an all sky survey of this type. 
The improvements could be even larger if two high density tracers with a significantly different
sensitivity to non-gaussianity can be identified and 
measured over a large volume. 

\end{abstract}


\maketitle

\setcounter{footnote}{0}

\section{Introduction}

One of the major unsolved mysteries in cosmology is the creation of structure in 
the universe. There are many competing theories that differ in their predictions, some of which are
accessible to cosmological observations today. 
Inflation \cite{Starobinsky:1979ty,1981PhRvD..23..347G,1982PhLB..108..389L,1982PhRvL..48.1220A},  the oldest and most successful of these theories, predicts that correlations of
initial structures are nearly scale invariant, adiabatic, and nearly gaussian \cite{1981JETPL..33..532M,1982PhLB..115..295H,1982PhRvL..49.1110G,1982PhLB..117..175S,1983PhRvD..28..679B}. 
However, ever since the first detection of cosmic microwave background anisotropies in COBE it has been recognized
that cosmic variance, a finite number of structures on the largest scales, provides a fundamental limitation 
to how well we can distinguish among competing models using two-point function correlations \cite{1993MNRAS.263L..17S,1993ApJ...417L...9S}. 
This is because the primordial density field is a single realization of a random process and only its two-point 
function is specified by the theoretical predictions. 
For many discriminating tests
such as the shape of the power spectrum on large scales, the cosmic variance limit is fundamental. 
As a consequence, many interesting questions open today may remain unanswered, such as the question of whether
the low quadrupole in CMB is due to manifestation of some new physics on very large scales or just a chance fluctuation \cite{2003ApJS..148..175S}. 

Our desire to discriminate among the models motivates searching for other probes, specially those where the cosmic variance limit can be avoided.
An important example where this limit can to some extent be avoided are primordial  
gravitational waves. This is due to the fact that there exists a polarization mode (so called B polarization) that only receives contributions 
from gravitational waves and not from the dominant scalar perturbations \cite{1997ApJ...482....6S,1997PhRvL..78.2054S,1997PhRvL..78.2058K}. 
In this case the limit is determined by instrumental and other irreducible noise rather than by the cosmic variance. 
Inflation allows for a detectable level of gravitational waves on the largest scales and as a result primordial gravitational waves have been 
declared as the smoking gun for inflation \cite{2001PhRvD..64l3522K}, based also on the fact that some competing 
models \cite{2001PhRvD..64l3522K,2003PhRvL..91p1301K,2002Sci...296.1436S} do not predict detectable 
levels of gravity wave signal and its detection would rule them out. The fact that cosmic variance can be avoided and that competing 
theories differ in their predictions are the primary motivators behind the interest in CMB polarization as the next experimental frontier.

One would like to identify other probes that can discriminate among the competing theories. 
One of these is non-gaussianity and
most of the models predict non-gaussianity of local type, $\Phi=\phi+f_{nl}\phi^2$, where $\Phi$ is the gravitational 
potential in the matter era and $\phi$ is the corresponding primordial gaussian case. 
In the simplest models of inflation primordial 
non-gaussianity is predicted to be very small, $f_{nl} \ll 1$. Alternatives to inflation, such as the 
new ekpyrotic/cyclic scenario 
naturally predict large non-gaussianity, $f_{nl} \gg 1$
\cite{2008PhRvL.100q1302B,2007JCAP...11..010C,Lehners:2007wc}, 
leading to non-gaussianity 
as the smoking gun for alternatives to the simplest models of inflation. 
However, 
all of the tests of non-gaussianity proposed so far suffer from cosmic variance limit. 
Until recently the most powerful method to 
place limits on $f_{nl}$ was based on bispectrum of cosmic microwave background (CMB), with the latest WMAP constraint 
$-9 < f_{nl}<111$ at 95\% confidence level \cite{2008arXiv0803.0547K}. The statistical power of this method is limited by the cosmic 
variance and with a better 
angular resolution one can sample more modes and thus improve the limits. For example, the limits
should be improved significantly with the higher angular resolution Planck satellite, where 
95\% interval on $f_{nl}$ of 30 is expected \cite{2008arXiv0803.4194C}, but this will be limited by astrophysical uncertainties such as the 
secondary anisotropies in CMB. Because the primordial CMB is damped on small scales it seems unlikely that the ultimate 
limits from CMB will be significantly improved over the current Planck predictions, but the promise of measuring high redshift 3-dimensional matter 
distribution with 21 cm transitions could improve the limits further, possibly to the level below unity \cite{2006PhRvL..97z1301C}.

An alternative approach using clustering of biased tracers of structure on very large scales has recently been proposed
\cite{2008PhRvD..77l3514D}. It was shown that non-gaussianity leads to a very unique scale dependence of the large scale bias,
one that increases strongly towards large scales, and whose amplitude scales with the bias of the tracer relative 
to the dark matter. One can therefore place the limits on $f_{nl}$ by comparing the scale dependence of the power spectrum of the 
biased tracer to the one expected in cosmological models under the assumption of a scale independent bias. 
Subsequent work explored further theoretical issues and prospects for the future \cite{2008ApJ...677L..77M,2008arXiv0805.3580S,2008arXiv0806.1046A}. 
A first application of this method, which we will call the power spectrum method, has recently been presented using 
large scale clustering of quasar and luminous red galaxies (LRG) galaxy data from Sloan Digital Sky Survey (SDSS) \cite{2008arXiv0805.3580S}. 
The result, $-29 < f_{nl}<69$ (95\% c.l.), is already better than the latest CMB constraints from WMAP, suggesting this is a competitive method compared to the 
bispectrum from CMB and should be pursued further. 
Projections for the future suggest that a survey of $10^6$ LRGs out to $z \sim 0.7$ over a quarter of the sky, such as
the proposed SDSS-III, will improve these limits by another factor of two \cite{2008PhRvD..77l3514D}. Further into the 
future, an all sky sample of highly biased tracers out to $ z\sim 2$ with $10^8$ galaxies, such as those contemplated for the dark energy 
mission \cite{2006astro.ph..9591A}, 
could reduce the width of the 95 \% confidence interval to the level of 5-10 \cite{2008arXiv0806.1061M,2008arXiv0806.1950C,2008arXiv0806.1046A}. 

While these predictions for the future are impressive, it would be great if one could 
improve them further to reach the realm of a guaranteed detection, 
which is at the level of $f_{nl} \sim 1$ \cite{2004PhR...402..103B}. 
There are two main obstacles to this. First, this method suffers significantly from cosmic variance: this is because
the signal is strongest on the largest scales where the cosmic variance, due to the finite number of realizations at a given scale 
within a given volume, makes the error on the power spectrum large. 
Relative error from each mode is of order unity, hence we can only detect the $f_{nl}$ signal  
if the relative change in power due to $f_{nl}$ is of order unity. Second limitation pointed out in \cite{2008arXiv0805.3580S} is that for current data
the effect of $f_{nl}$ is correlated with other cosmological parameters, such as matter density and primordial slope of the power spectrum, 
which also change the shape of the power spectrum on large scales. 
This is because over the limited range of scales accessible these parameters can partially mimic the effect of $f_{nl}$. It is
particularly important 
for two-dimensional projections, such as the quasar and LRG samples based on photometric data used in \cite{2008arXiv0805.3580S}. 
For a given multipole moment $l$ the 
k-space projection window is broad and the effect of $f_{nl}$ extends to fairly high $l$, but it does not have a very characteristic 
scale dependence that would allow one to separate it from the other parameters. 

Above discussion suggests, firstly, that
methods that can avoid the cosmic variance limit are particularly useful 
for future progress in discriminating between competing theories and, secondly,  that non-gaussianity is a key probe 
that can differentiate among them. 
In this paper we propose a method to probe non-gaussianity from large scale structure that circumvents the cosmic variance limit and 
also eliminates the problem of its degeneracy with other cosmological parameters. Instead of measuring the power spectrum of a tracer 
and comparing it to predictions based on cosmological models we propose to compare directly the density field of a biased tracer to the 
one of an unbiased tracer (or, more generally, to a tracer with a different bias, which may include less biased or even anti-biased tracers). 
The relative bias of the two tracers is scale independent in the absence of non-gaussianity, but picks up a scale dependence in its presence. 
The main advantage is that tracers are generally biased, but not stochastic, on very large scales. This means 
that the precision with which the relative bias between the two tracers can be determined 
is only limited by the noise, which is given by the Poisson sampling of the field, and not by the cosmic variance. 
Moreover, because we are directly comparing two density fields any scale dependence of relative bias can only be caused by non-gaussianity and 
not by other cosmological parameters, which cannot affect the amplitude ratio even if they can affect the power spectrum. 
Hence, there is no degeneracy between $f_{nl}$ and other cosmological parameters with this method. 
In the next section we quantify these statements, computing the error predictions for both methods and 
comparing them to each other.  We then apply it to 
some representative examples of present and future data sets, showing the improvements that can be expected. 
We finish by highlighting the requirements needed for the future surveys 
to maximize the power of this method. 

\section{Error Analysis} 

Let's assume we measure two tracers of matter density field, $\delta_1$ and $\delta_2$. They are both biased tracers of the underlying 
matter density field $\delta$, $\delta_i=b_i\delta$, where $b_1$, $b_2$ is the large scale bias of two tracers. We can introduce relative bias $\alpha=b_1/b_2$ and the corresponding covariance matrix elements in the Fourier domain are 
$C_{22}=<\delta_2^2>=(P_2+\bar{n}_2^{-1})/V$, $C_{12}=<\delta_1\delta_2>=r\alpha P_2/V$ and $C_{11}=<\delta_1^2>=(\alpha^2P_2+\bar{n}_1^{-1})/V$, where $r$ is the cross-correlation coefficient 
between the two fields, $P_2$ is the power spectrum of second tracer and $V$ is the volume element (which drops out from the final expressions). 

We want to compare the errors on non-gaussianity parameter $f_{nl}$ as extracted from the power spectrum analysis to the one from the analysis of 
the relative amplitude of the two tracers. To answer this we need to compute the errors of 
the corresponding parameters, $P_2$ and $\alpha$, and their dependence on $f_{nl}$. We will first compute the 
errors for these two parameters and then combine with their dependence on $f_{nl}$ to derive the final error predictions on $f_{nl}$. 
Fisher matrix plays a key role in describing the ability of a survey to constrain parameters such as $f_{nl}$. Its inverse 
gives the expected covariance matrix of parameters one wishes to estimate. 
It is defined as 
\begin{equation}
F_{\lambda \lambda'}={1 \over 2} tr[C_{,\lambda}C^{-1}C_{,\lambda'}C^{-1}],
\label{fisher}
\end{equation}
where $C$ is the covariance matrix of the data defined above and $\lambda$ is the set of parameters one is estimating. 

Let us begin with the estimated error of the power spectrum $P_2$ for a single mode. 
Applying equation \ref{fisher} we find 
\begin{equation}
F_{P_2P_2}={(\alpha^2X_2+X_1)^2+2(1-r^2)\alpha^2 \left[\alpha^2(1-r^2)+X_2+X_1(1+X_2) \right] \over 2P_2^2[\alpha^2(1-r^2)+\alpha^2X_2+X_1+X_1X_2]^2},
\end{equation}
where we defined $X_i=(\bar{n}_iP_2)^{-1}$. If $X_i$ is larger than unity then we are in the Poisson noise 
dominated limit 
and Poisson noise dominates over the sampling variance. In the opposite limit 
$X_i\ll 1$ sampling variance dominates the error and the above expression reduces to $F_{P_2P_2}^{-1}=\sigma_{P_2}^2=2P_2^2$ if $r=1$, 
which is the usual sampling variance error for one mode. Hence the relative error from a single mode is limited from below to $2^{1/2}$
and this irreducible error is called the cosmic variance. 

We can also apply the above expressions to compute the diagonal component of the Fisher matrix for relative amplitude $\alpha$,
\begin{equation}
F_{\alpha \alpha}={\alpha^2 X_2(1 +2X_2)+r^2X_1(1+X_2)+\alpha^2(1-r^2)(2-r^2+3X_2) \over [\alpha^2(1-r^2)+\alpha^2X_2+X_1+X_1X_2]^2},
\end{equation}
which in the limit $X_1 \ll 1$, $X_2\ll 1$, $1-r^2 \ll 1$ becomes
\begin{equation}
F_{\alpha \alpha}^{-1}=\sigma_{\alpha}^{2}=\alpha^2 X_2 +X_1+\alpha^2(1-r^2). 
\end{equation}
Thus the error on $\alpha$ from a single mode can be much less than unity if there 
is little stochasticity ($r \sim 1$) and the field is oversampled, $X_1, X_2 \ll 1$. 
Therefore, there is no fundamental cosmic variance limit here and if $\alpha$ depends on $f_{nl}$ then there is no 
cosmic variance limit on the latter either. 

To connect these expressions to the precision with which one can determine $f_{nl}$ we need to look at the dependence of $P_2$ and 
$\alpha$ on $f_{nl}$. 
For a given tracer with large scale bias $b$ we have \cite{2008PhRvD..77l3514D,2008ApJ...677L..77M,2008arXiv0805.3580S,2008arXiv0806.1046A}
\begin{equation}
P(k)=(b+\Delta b(k)f_{nl})^2P_{dm}(k),
\end{equation}
where $P_{dm}(k)$ is the dark matter power spectrum and 
\begin{equation}
\Delta b(k)=2(b-p)\delta_c {\phi \over\delta}={3(b-p)\delta_c\Omega_m H_0^2 \over c^2k^2T(k)D(z)},
\end{equation}
where $\delta_c=1.686$ is the spherical collapse linear over-density, $\phi$ is the primordial potential in matter domination, 
$\delta$ is the matter overdensity, $H_0$ is the Hubble parameter, $c$ is the speed of light, $T(k)$ is the transfer 
function and $D(z)$ is the linear growth rate normalized to $(1+z)^{-1}$ for $z\gg 1$. For tracers whose selection only depends on halo mass
we have $p=1$, but this can be larger if the tracer also depends on other properties such as the merging history, where 
for recent mergers one has $p \sim 1.6$ \cite{2008arXiv0805.3580S}. Here we will use $p=1$ in the following, but one should keep in mind
that this is not always appropriate. 

Fisher matrix for $f_{nl}$ is 
\begin{equation}
F_{f_{nl}f_{nl}}=\sum_{\lambda,\lambda'} F_{\lambda \lambda'}{\partial \lambda \over \partial f_{nl}}{\partial \lambda' \over \partial f_{nl}}. 
\end{equation}
Let us continue to treat the two ways of estimating $f_{nl}$ separately. From the power spectrum estimator for a single mode we find 
$F_{f_{nl}f_{nl}}(P_2)=F_{P_2P_2}(2\Delta b(k)/bP_2)^2$, which in the sampling variance limit gives
\begin{equation}
\sigma_{f_{nl}}(P_2)={b \over 2^{1/2}\Delta b(k)}. 
\end{equation}
Here we have assumed the non-gaussian correction is small to simplify the expressions. We have also assumed that the 
two samples have been combined optimally into a single sample with an overall bias $b$ and non-gaussianity
dependence $\Delta b$, 
the details of which will depend on their number density and bias. For example, 
the simplest case is if $b_2=1$ in which case we can assume this tracer contains no information on $f_{nl}$ and the overall sample is just the biased sample with $b=b_1$ and $\Delta b= \Delta b_1$. 
The overall error is
obtained by integrating over all modes
\begin{equation}
\sigma_{f_{nl}}^{-2}(P_2)= { V \over \pi^2}\int_{k_{min}}^{k_{max}} (\Delta b(k) /b)^2k^2dk, 
\end{equation}
where we assumed the sampling variance limit ignoring the Poisson term. 
Here $V$ is the volume of the survey, $k_{min} \sim 2\pi/V^{1/3}$ is the smallest wavector accessible in such a survey and $k_{max}$ is the 
largest wavevector for which we can assume validity of this expression. 
Its value is often not very important since most of the sensitivity comes from large scales, ie small wavevectors. 
Note that we have ignored any degeneracies between $f_{nl}$ and other cosmological parameters in this expression, so it is bound to 
be overly optimistic, although for anything more quantitative one would need to do the actual analysis with a given survey geometry and 
choosing the cosmological parameters one is varying in the analysis. 

The Fisher matrix for  $f_{nl}$ from the relative amplitude of two tracers for a single mode is given by
\begin{equation}
\sigma_{f_{nl}}(\alpha)=\left(X_2+X_1\alpha^{-2}+1-r^2\right)^{1/2}\left({\Delta b_1 \over b_1}-{\Delta b_2 \over b_2}\right)^{-1}. 
\end{equation}
Integrating over all the modes gives
\begin{equation}
\sigma_{f_{nl}}^{-2}(\alpha)={ V \over 2\pi^2}\int_{k_{min}}^{k_{max}}F_{\alpha \alpha}\alpha^2\left({\Delta b_1(k) 
\over b_1}-{\Delta b_2(k) \over b_2}\right)^2k^2dk. 
\end{equation}
Here we have assumed that $\alpha=b_1/b_2$ is known perfectly from the smaller scales where $f_{nl}$ effects are negligible but linear 
bias still applies, so that information on $\alpha$ from large scales 
is used for the estimation of $f_{nl}$ rather than the relative bias itself. 

It is useful to compare the ratio of expected errors for a single mode in the sampling variance limit assuming $r=1$. We find
\begin{equation}
{\sigma_{f_{nl}}^2(\alpha) \over \sigma_{f_{nl}}^2(P_2)}=2(X_2+X_1/\alpha^2)\left[{\Delta b /b \over \Delta b_1/b_1-\Delta b_2/b_2}\right]^2.  
\end{equation}
If we assume we have two tracers, one unbiased, $b_2=1$,  which contains no information on $f_{nl}$, and one biased with $b_1=b$
this simplifies to 
\begin{equation}
{\sigma_{f_{nl}}(\alpha) \over \sigma_{f_{nl}}(P_2)}= \sqrt{2(X_2+X_1/\alpha^2)}=\sqrt{2[(\bar{n}_2P_2)^{-1}+(\bar{n}_1P_1)^{-1}]}.
\label{sn}
\end{equation}
%For the latter expression we used
%that $X_2=(\bar{n}_2P_2)^{-1}$ is the noise to signal power ratio of tracer 2 and $X_1/\alpha^2$ is the noise to signal power ratio of tracer 1, $X_1/\alpha^2=(\bar{n}_1P_1)^{-1}$. 
The dependence on $\Delta b$ and $b$ has dropped out from this expression.
Finally, if one assumes the abundance of unbiased tracers
is much larger than that of biased tracers we can drop $X_2$ term and we 
conclude that the signal to noise improvement per mode is simply given by the signal to noise power ratio of the biased tracer times root two, 
$\sigma_{f_{nl}}(\alpha) / \sigma_{f_{nl}}(P_2)=(2/\bar{n}_1P_1)^{1/2}$, where $P_1$ is the power spectrum of the biased tracer. 

\section{Discussion} 

Let us apply the above derived relations to some examples of interest. 
SDSS-III plans to do a spectroscopic survey of about $1.5\times 10^6$ luminous red galaxies (LRGs) out to $z \sim 0.7$ over a quarter of 
the sky. The expected number density is about $3 \times 10^{-4}{\rm (h/Mpc)^3}$ and average bias about $b=1.7$ \cite{2007MNRAS.378..852P}, so $P_1(k=0.01{\rm h/Mpc})=0.8 \times 
10^5{\rm (Mpc/h)^3}$ 
and we may expect about a factor of 3 error improvement if the unbiased tracers are available and their shot noise can be neglected. This latter 
assumption is by no means easy to achieve, since lower bias galaxies live in lower mass halos,
are typically fainter and SDSS-III will not target them spectroscopically. One way to pursue is using photometric samples, which is discussed 
further below. Another is to split the LRG sample into two luminosity bins, where
we may expect for the brighter bin $\bar{n}_1=10^{-4}{\rm (h/Mpc)^3}$ and $b \sim 2$, so $P_1(k=0.01{\rm h/Mpc})=1.2 \times 10^5{\rm (Mpc/h)^3}$ at $z \sim 0.5$,
while for the fainter bin we may expect $n_2=2\times 10^{-4}{\rm (h/Mpc)^3}$ and $b=1.4$, in which case 
$d\alpha/d f_{nl}$ is reduced by more than a factor of two and the noise is increased since $X_1=X_2=0.1$. In this case this method gives a 
comparable signal to noise to the power spectrum method. This may still be advantageous, since there is no degeneracy with other 
cosmological parameters and one can extend the analysis to small scales. 

Looking further into the future, there are a number of planned spectroscopic surveys that 
will measure a high number of redshifts to higher redshifts \cite{2008ExA...tmp...12C,2006astro.ph..6104P}. In this case
one may expect larger improvements, specially 
if all of the most biased halos can be identified. For example, 
at $ z \sim 1.8$ we may have $\bar{n}_1 \sim 3 \times 10^{-3}{\rm (h/Mpc)^3}$ halos with average bias $b=2$ and $P_1(k=0.01{\rm h/Mpc}) \sim 3 \times 
10^4{\rm (Mpc/h)^3}$ and most of these should be detectable with the above mentioned surveys. 
Hence in this case the improvement 
can be a factor of 7 over the power spectrum method. 
Note that with the power spectrum method such an improvement would require a 50-fold increase in the volume of the survey. 
This is a particularly exciting prospect since the predicted 95\% confidence level interval for a full 
sky survey to $z \sim 2$ is of order $\Delta f_{nl} \sim 5-10$ 
\cite{2008arXiv0806.1061M,2008arXiv0806.1950C,2008arXiv0806.1046A} and with the additional factor of 7 reduction of error we may be 
able achieve $\Delta f_{nl} \sim 1$ (95\% c.l. interval), at which point we enter into regime of a guaranteed detection \cite{Bartolo:2003bz}. 

The benefits may be even greater for photometric surveys, where no spectroscopy is needed and hence can be done with a significantly smaller
investment. 
On very large scales the advantages of a spectroscopic survey become less important, 
specially if the photometric redshift error is well below the largest scale of the survey and if, when
several redshift slices are available, the 
cross-correlation information between photometric samples is included in analysis. 
Indeed, currently the strongest constraints come from the power spectrum method using 
LRG photometric sample and 
quasar photometric sample, both from SDSS \cite{2008arXiv0805.3580S}. 
As mentioned above, the main limitation of photometric samples is that the projection along the line of sight causes the non-gaussianity effect to extend 
over a wide range of multipole moments, making the effect difficult to distinguish from other cosmological parameters using
the power spectrum method \cite{2008arXiv0805.3580S}.  This problem is eliminated using the method proposed here. 

Another potential advantage is that it is easier to obtain a high density sample of unbiased tracers at the same redshift using photometric data. 
For the LRG sample that goes to $z \sim 0.7$ the corresponding photometric low bias sample already exists
from SDSS
with well calibrated and reasonably accurate 
photometric redshifts and high number densities at the level of $\bar{n}=10^{7}$ (number of galaxies 
per steradian) \cite{2008MNRAS.386..781M}. For photometric LRGs with their 
number density of $\bar{n}=5\times 10^{5}$ and power $C_l=1.5\times 10^{-4}$ at $l=6$ \cite{2007MNRAS.378..852P}
we find a possible improvement of a factor of 5 over the current LRG photometric constraints of $\Delta f_{nl} \sim 200$ for 95\% confidence interval. 
A potential challenge of this approach is matching the redshift distribution of the two samples, since any mismatch gives rise to less than perfect correlations. 
This can be achieved by constructing a set of weights for each galaxy that takes into account its redshift error distribution. 
It is also possible to combine spectroscopic and photometric samples. For example, one can use a high bias spectroscopic sample, such as the
above mentioned LRG spectroscopic survey from SDSS-III,
and apply the radial weights to reproduce the redshift distribution of a photometric sample. Finally, we should also note that the ultimate unbiased
tracer insensitive to $f_{nl}$ to correlate against 
can be the dark matter itself, as measured from weak lensing, although in this case the radial window is rather 
broad \cite{2004MNRAS.350.1445P}. 
Future photometric surveys may allow for much more significant improvements. For example, PAU survey \cite{2008arXiv0807.0535B} plans to measure 
galaxies with a number density of $\bar{n}=10^{-3}{\rm (h/Mpc)^3}$ out to $z \sim 0.9$, so here both the biased and unbiased samples
will be available. The expected accuracy of photometric redshifts, $dz \sim 0.003(1+z)$, is sufficiently small that it can be ignored for the large scales 
of interest here. The expected error improvement will be about a factor of 3 over the power spectrum method and the predicted 95\% c.l. interval
of $\Delta f_{nl} \sim 5$ is approaching the levels of a guaranteed detection. 

These predictions are based on the assumption that there is no stochasticity in the tracers, ie we assumed $r=1$ in the analysis. 
It is well known that halos and galaxies are not perfect tracers of the density field and on smaller scales stochasticity develops. 
On small scales and/or for high density samples this can become the limiting factor: as shown in equation \ref{sn} this happens when $1-r^2$ exceeds 
$(\bar{n}P)^{-1}$ and since we need the latter to be well below unity for the method to be superior over the power spectrum method this is 
a potential issue that needs further investigation. 
The large scale cross-correlation coefficient between different tracers has not been studied in much detail with numerical simulations. In \cite{2004MNRAS.355..129S}
the related cross-correlation between final dark matter and halos has been investigated, but the small box size used in the analysis 
makes the results noisy. Still, the results suggest that the cross-correlation coefficient is indeed small on very large scales, with 
$1-r^2<0.1$ for $k<0.05{\rm h/Mpc}$ and likely to be even smaller on larger scales. 
An analysis with a much larger volume is in progress and preliminary results in configuration space suggest
that $1-r^2 \sim 0.1$ at 5 Mpc/h and $1-r^2 \sim 0.01$ for $r>50$ Mpc/h \cite{smithseljak}. 
This issue clearly requires further attention and more detailed studies using very large volume simulations with biased tracers will be needed to address it
in detail to see if 
cross-correlation coefficient is sufficiently close to unity on very large scales to allow the several fold reduction of error promised
by this method. 

While we have focused on the biased tracers of large scale structure, it is also possible to 
look for the signature using the anti-biased tracers. For low mass halos the bias is asymptotically approaching $b \sim 0.7$ \cite{2004MNRAS.355..129S,2005MNRAS.363L..66G} and for this population the effect of $f_{nl}$ has the opposite sign from the biased population with an $f_{nl}$
dependence that is about a factor of 3-4 smaller than for $b=2$ population assuming $p=1$. 
In addition, such an anti-biased tracer has about a factor of 10 smaller amount of power. 
These two disadvantages can in principle be offset by the higher number density, which can be
orders of magnitude larger for $b \sim 0.7$ population relative to the $b \sim 2$ population. 
We can also use these antibiased tracers instead of the unbiased tracers and comparing them to the biased tracers, 
which improves the sensitivity to 
$f_{nl}$ and reduces the shot noise relative to the unbiased tracer we assumed in our analysis. 

More generally, recent studies suggest that bias can depend on quantities other than the 
halo mass \cite{2005MNRAS.363L..66G,2007MNRAS.377L...5G}, and similarly one can also expect secondary parameters that may enhance or
suppress the sensitivity to $f_{nl}$.
Indeed, an extended Press-Schechter analysis suggests one 
such second parameter that suppresses the sensitivity to $f_{nl}$ of highly biased tracers is the recent merger activity \cite{2008arXiv0805.3580S}. 
Observationally identifiable candidates that satisfy this criterion may for example be quasars.
The challenge for the future is to find two tracers, one that exhibits a significant dependence on $f_{nl}$ and one that does not, both of which 
come with a sufficiently high number density 
to reduce the Poisson noise and both of which can be observationally identified within existing and future surveys. 
Given the potentially huge payoff of finding such tracers for the questions of interest to the fundamental theories of the universe 
this is a challenge that is worth exploring further 
both with observations as well as with numerical simulations and analytic methods. 

I thank Nikhil Padmanabhan and An\v ze Slosar for useful comments on the draft. 
U.S. is supported by the
Packard Foundation and 
Swiss National Foundation
under contract 200021-116696/1.

\bibliography{cosmo,cosmo_preprints}
\end{document}